\documentclass{ecai} 

\usepackage{latexsym}
\usepackage{amssymb}
\usepackage{amsmath}
\usepackage{amsthm}
\usepackage{booktabs}
\usepackage{enumitem}
\usepackage{graphicx}
\usepackage{color}
\usepackage{adjustbox}
\usepackage{subcaption}
\usepackage{multirow}

\newcommand{\BibTeX}{B\kern-.05em{\sc i\kern-.025em b}\kern-.08em\TeX}

\begin{document}


\begin{frontmatter}


\paperid{5326} 


\title{Robust Coordination under Misaligned Communication \\
via Power Regularization}


\author[A]{\fnms{Nancirose}~\snm{Piazza}\orcid{}}
\author[A]{\fnms{Amirhossein}~\snm{Karimi}\orcid{}\thanks{Corresponding Author. Email: akari9@unh.newhaven.edu.}\footnote{Equal contribution.}}
\author[B]{\fnms{Behnia}~\snm{Soleymani}\orcid{}\footnotemark}
\author[A]{\fnms{Vahid}~\snm{Behzadan}\orcid{}}
\author[C]{\fnms{Stefan}~\snm{Sarkadi}\orcid{}}

\address[A]{SAIL Lab, University of New Haven, United States}
\address[B]{University of Tehran, Iran}
\address[C]{King's College London, United Kingdom}


\begin{abstract}
Effective communication in Multi-Agent Reinforcement Learning (MARL) can significantly enhance coordination and collaborative performance in complex and partially observable environments. However, reliance on communication can also introduce vulnerabilities when agents are misaligned, potentially leading to adversarial interactions that exploit implicit assumptions of cooperative intent. Prior work has addressed adversarial behavior through power regularization through controlling the influence one agent exerts over another, but has largely overlooked the role of communication in these dynamics. This paper introduces Communicative Power Regularization (CPR), extending power regularization specifically to communication channels. By explicitly quantifying and constraining agents' communicative influence during training, CPR actively mitigates vulnerabilities arising from misaligned or adversarial communications. Evaluations across benchmark environments Red-Door-Blue-Door, Predator-Prey, and Grid Coverage demonstrate that our approach significantly enhances robustness to adversarial communication while preserving cooperative performance, offering a practical framework for secure and resilient cooperative MARL systems.
\end{abstract}

\end{frontmatter}


\section{Introduction}

Coordination in Multi-Agent Reinforcement Learning (Comm-MARL) involves synchronizing actions among agents that operate in a shared environment but possess individual objectives, often aiming to prevent detrimental outcomes. Within this framework, communication is understood through the perspectives of information theory and control, defined as the exchange of information between agents via an established channel, typically employed to facilitate coordination. In contrast, Cooperative Multi-Agent Reinforcement Learning (CoMARL) generally emphasizes parameter-sharing, optimizing team training efficiency, and developing cooperative mechanisms to address collective challenges. While many CoMARL algorithms leverage parameter-sharing for training efficiency, the resulting joint policies can lead to undesirable behaviors. These include individual agent free-riding \citep{ueshima2023deconstructingcooperationostracismmultiagent}, over-reliance on or learning of irrelevant features and conventions \citep{köster2020modelfreeconventionsmultiagentreinforcement}, and insufficient experience to respond effectively to actions from misaligned agents, such as those encountered in adversarial scenarios.

Objective misalignment characterizes multi-agent systems where agents are non-cooperative, potentially indifferent to their impact on others, and pursue self-interested goals. In settings with misaligned agents, public communication channels are vulnerable to misuse or sabotage, particularly against cooperative agents trained to rely on signaling through these channels. A critical requirement for multi-agent systems (MAS) communication, especially when deploying teams of cooperative, autonomous agents in real-world scenarios, is resilience to such objective misalignment. This resilience requires evaluating objective misalignment under conditions that are not typically addressed by standard CoMARL algorithms, while taking into account both team and individual contexts.

Communication remains a significant research area in collaborative and cooperative MARL \citep{oroojlooyjadid2021reviewcooperativemultiagentdeep}. Much of the MARL literature models communication using dedicated protocol controllers, like those in CommNet \citep{10.5555/3157096.3157348} and IC3Net \citep{singh2018learningcommunicatescalemultiagent}, which manage information propagation and evaluate the joint policy. Typically, in settings where agents learn to communicate, informed senders learn to encode useful messages, while uninformed receivers learn to condition their actions on received messages. When agents concurrently learn communication and environment policies, they might intrinsically develop uncontrolled regularization against misaligned communication, whether (1) these arise from co-learning errors by other agents or (2) are intentionally generated by agents with differing objectives. However, self-learned communication may not be beneficial in all scenarios, potentially creating vulnerabilities that adversaries can exploit through adversarial communication directed at naive learned behaviors.

We define misaligned communication as any message transmitted over an established channel that negatively affects a recipient agent's performance, irrespective of adversarial intent or the presence of an explicit adversary. In the context of MARL communication, fostering resilience to misaligned communication within cooperative settings should encourage cooperative agents to adapt, learning policies robust enough for competitive or mixed settings. This differs from adversarial attacks, which involve an explicit adversary with zero-sum objectives targeting a specific model or team, exploiting system components to inject malicious payloads. Research on adversarial attacks targeting multi-agent communication demonstrates that compromising the communication channel can severely degrade the performance of communication-dependent agents \citep{tu2021adversarialattacksmultiagentcommunication}, a challenge compounded in multi-channel communication scenarios \citep{dong2022multiagentadversarialattacksmultichannel}, further complicating the training of large-scale communicating MARL systems.

Power, representing one agent's influence over another's decision-making, can be modeled by relating parts of an agent's expected utility exploited by the actions of other agents. Incorporating formal measures of power, based on expected utility, into the training process offers a potential avenue for enhancing policy robustness, despite the inherent complexity. While agents typically do not explicitly optimize for power over others, reconstructing utility functions as additive components representing different utility types could provide greater control over agent behaviors. Analogous to intrinsic rewards in MARL \citep{NEURIPS2019_07a9d3fe}, reconstructing forward optima might offer a viable alternative to attempting to analyze existing black-box policies.

While communication in CoMARL mainly serves to coordinate agents, it is important to consider its value outside of purely cooperative scenarios that also need careful evaluation. While highly beneficial in cooperative settings, excessive reliance on communicated messages can make policies vulnerable to exploitation and opportunistic behaviors. We propose power regularization as a mechanism for designers to manage behaviors exhibiting power dependency, making policies more suitable for competitive and mixed environments. Power regularization over communication involves trade-offs, as it encourages agents to learn alternative behaviors, potentially at the cost of optimal expected utility. Furthermore, the feasibility of these alternatives depends on the range of acceptable strategies within the environment. Designers must ultimately assess whether policies exhibiting learned resilience to specific forms of misaligned communication achieve acceptable performance levels.

In this paper, we employ power regularization over communication as a method to develop communication policies that balance reliance on influential agents with the value of individual agent autonomy.

Our contributions are as follows:
We introduce power regularization as a technique to mitigate the adverse effects of misaligned communication in CoMARL. By quantifying power expressed through communication, designers gain a tool to control power dynamics within learned policies.

We evaluate the effectiveness of power regularization over communication in three benchmark environments, Grid Coverage, Predator-Prey, and Red-Door-Blue-Door, demonstrating its ability to reduce the impact of misaligned communication compared to standard cooperative baselines in different settings.

This paper is organized as follows: We begin with background information covering related work, communication in CoMARL, and the concept of measuring power as part of the optimization criterion. Subsequently, we introduce power regularization over communication as a specific application of the power formulation and detail our experiments in the Grid Coverage, Predator-Prey, and Red-Door-Blue-Door environments. Finally, we conclude by discussing the limitations of our approach and suggesting directions for future research.

\section{Related Work}

Adversarial communication in MARL settings is often highlighted by its emergence in non-cooperative settings \citep{blumenkamp2020emergenceadversarialcommunicationmultiagent}, which may be the product of misaligned agents. Adversarial attacks in MARL settings are diverse in their methodology, ranging from sparse targeted attacks \citep{hu2022sparseadversarialattackmultiagent} to attacks that exploit vulnerabilities in mechanism design, such as consensus-based mechanisms \citep{figura2021adversarialattacksconsensusbasedmultiagent} and adversarial minority influence \citep{li2024attackingcooperativemultiagentreinforcement}. Adversarial training, an approach to mitigating against adversarial interests, is an umbrella-term for incorporating adversarial interactions into training for hardening and better resilience against adversarial opponents. In support, there are works \citep{lin2020robustnesscooperativemultiagentreinforcement, guo2022comprehensivetestingrobustnesscooperative} on the robustness of CoMARL. There are diverse defenses against adversarial communication, including works that consider test-time settings with theory of mind inspired mechanisms \citep{piazza2023theorymindapproachtesttime}. In our work, adversarial training is used to address misaligned communication. Many CoMARL works that address credit assignment between global reward and local reward can be viewed as a means for regularizing agent behaviors and dynamics. For example, a reward-shaping mechanism \citep{ibrahim2020reward} was proposed to portion out the team reward based on individual contributions in order to address free-riders. Foerster et al. proposed COMA, Counterfactual Multi Agent Policy Gradients \citep{Foerster_Farquhar_Afouras_Nardelli_Whiteson_2018}, which marginalizes out single agent actions and also addresses credit assignment with the incentive that agents will maximize their contribution to the global reward. The motivation behind COMA is complementary in the sense that it quantifies how much influence an individual agent’s action has on the joint action, whereas this work quantifies how much influence other agents' actions have upon an individual agent’s policy. Additionally, there is investigative work on the causal relationship among agents in MARL \citep{pmlr-v97-jaques19a} through counterfactual reasoning, which, while promoted for more efficient and meaningful communication and coordination, can be a direction that quantifies the contribution of other agents to a self-agent’s return. Some other related works on explicit regularization in MARL originate from maximum-entropy MARL, which reconstructs the return as the reward and the entropy of the policy distribution, weighed by a temperature parameter. An example would be FOP \citep{pmlr-v139-zhang21m}, an actor-critic method that factorizes the optimal joint policy from maximum-entropy MARL. The existing work on quantifying power \citep{li2024benefitspowerregularizationcooperative} in MARL studies adversarial power, power associated with an adversarial opponent. The authors discuss various fine-tune parameters for implementing power and measuring power in multi-opponent settings. This work investigates power in settings with communication.

\section{Preliminaries}

\subsection{Communicative MARL}

Multi-Agent Reinforcement Learning (MARL) provides a framework for sequential decision-making problems involving multiple interacting agents. Cooperative MARL scenarios are often formalized as Decentralized Partially Observable Markov Decision Processes (Dec-POMDPs), representable by the tuple:
\[
\langle \mathcal{N}, \mathcal{S}, \{\mathcal{A}^i\}_{i \in \mathcal{N}}, \{\mathcal{O}^i\}_{i \in \mathcal{N}}, P, \{\mathcal{R}^i\}_{i \in \mathcal{N}}, \gamma \rangle
\]
Here, $\mathcal{N}$ denotes the set of $N$ agents, $\mathcal{S}$ is the global state space, $\mathcal{A}^i$ is the action space for agent $i$, and $\mathcal{O}^i$ is its observation space. The function $P(s'|s, \mathbf{a})$ defines the state transition dynamics, where $\mathbf{a} = \{a^i\}_{i \in \mathcal{N}}$ is the joint action assembled from individual agent actions sampled from their policies, $a^i \sim \pi^i(\cdot|o^i)$. Each agent $i$ receives a local observation $o^i$ and a reward $r^i = \mathcal{R}^i(s, \mathbf{a})$. In fully cooperative settings, agents typically share a common team reward, $r^i = R(s, \mathbf{a})$. The collective goal is to learn policies that maximize the expected discounted return
\(
G = \sum_{t=0}^T \gamma^t r_t
\)
where $\gamma \in [0, 1]$ is the discount factor.

A prevalent paradigm for training MARL agents is Centralized Training with Decentralized Execution (CTDE). During the training phase, CTDE algorithms utilize global information, such as the full state or the actions of all agents, to facilitate learning. However, during execution, each agent must operate solely based on its local observation history. For instance, Value Decomposition Networks (VDN) \citep{sunehag2017valuedecompositionnetworkscooperativemultiagent} learn decentralized policies by decomposing the global team Q-function $Q_{tot}(s, a)$ into a sum of individual agent Q-functions $Q^i(o^i, a^i)$:
\[
Q_{tot}(s, a) = \sum_{i \in \mathcal{N}} Q^i(o^i, a^i)
\]
Such methods often employ parameter sharing across agent networks to improve learning efficiency.

To further enhance coordination, particularly under partial observability, agents can utilize explicit communication. Communication channels allow agents to exchange information directly. Typically, agent $j$ generates a message $m^j$ based on its internal state or history $h^j$ via a learned communication policy $m^j \sim C^j(\cdot|h^j)$. Agent $i$'s effective input $s^i_{input}$ for decision-making can then incorporate received messages $m^{-i}$ from other agents alongside its own local observation $s^i_{input} = f(o^i, m^{-i})$. Various architectures facilitate this exchange. For example, CommNet employs shared network modules and averages messages over multiple communication rounds before action selection. Other approaches, like IC3Net, focus on learning selective communication strategies through gating mechanisms, often optimizing these alongside individual policies using individual rewards. The structure of communication, dictating which agents can exchange messages, is frequently modeled using graph representations.

\subsection{Implicit Communication via Graph Neural Networks} 

Communication can also be learned implicitly through structured feature aggregation. As explored in \citep{li2020graphneuralnetworksdecentralized} and utilized in our experiments, Graph Neural Networks (GNNs) offer a powerful mechanism for this. Instead of learning explicit messages, agents first process their local observations $o^i_t$ to generate feature embeddings $x^i_t \in \mathbb{R}^F$. These are stacked into a matrix $X_t = [x_t^1, \dots, x_t^N]^T \in \mathbb{R}^{N \times F}$. The GNN operates over a dynamic communication graph represented by an adjacency matrix (or Graph Shift Operator) $S_t \in \mathbb{R}^{N \times N}$, where $[S_t]_{ij} = 1$ if agent $j$ can transmit to agent $i$ at time $t$. (typically based on proximity).

The core mechanism is a graph convolution layer. For a single layer GNN transforming input features $X_{in} \in \mathbb{R}^{N \times F}$ to output features $X_{out} \in \mathbb{R}^{N \times G}$, the operation can be defined as:
\[
X_{out} = \sigma \left( \sum_{k=0}^{K-1} S_t^k X_{in} A_k \right)
\]
Here, $S_t^k X_{in}$ represents features aggregated from the $k$-hop neighborhood, effectively requiring $k$ rounds of message passing or communication exchanges. $K$ is the maximum communication hop count (filter size) defining the spatial receptive field of the graph convolution. $A_k \in \mathbb{R}^{F \times G}$ are learnable weight matrices specific to hop $k$, transforming and combining features across hops and dimensions. $\sigma(\cdot)$ is a non-linear activation function applied element-wise. For multi-layer GNNs ($L$ layers), this operation is cascaded:
\[
X_l = \sigma[\mathcal{A}_l(X_{l-1}; S_t)] \quad \text{for } l=1,\dots,L
\]
where $X_0$ is the input to the first layer and $\mathcal{A}_l$ denotes the graph convolution operation at layer $l$.

This GNN architecture learns what information is relevant to share and aggregate from the local neighborhood defined by $S_t$ and $K$. The resulting aggregated feature vector for agent $i$, denoted $[X_L]_i \in \mathbb{R}^{G_L}$ (the $i$-th row of the final layer's output), captures context from its communicative neighbors and serves as the input to its decentralized policy network $\pi^i(a^i | [X_L]_i)$.


\subsection{Power}

The concept of power refers to the influence one agent has over another agent's decision-making and utility. In shared environments, power dynamics play a critical role in determining how agents interact and coordinate. Li and Dennis \citep{li2024benefitspowerregularizationcooperative} introduced power as a formal measure, redefining the optimization criterion as a combination of expected task return and power utility. By incorporating power regularization into the training process, agents can learn policies that are more resilient to states where power imbalances make them vulnerable. Specifically, power quantifies the expected difference between the current joint policy and a hypothetical joint policy where other agents take adversarial actions over $k$ steps.

Power is closely related to the concept of game security, which evaluates the expected return when facing adversarial opponents. In sequential games, the minimax strategy is often used to select the best action among the worst possible outcomes. When power dynamics naturally emerge or are necessary for completing team tasks, power regularization provides designers with a mechanism to control the autonomous behaviors of agents, ensuring they remain robust to adversarial influences.

The original formulation of power by Li and Dennis, referred to as standard power in this paper, estimates the influence agent $j$ has over agent $i$ as follows:

\[
\rho_{i:j}^\text{standard}(\pi^i, \pi^j, s) = Q_i^{\pi^i, \pi^j}(s, a^i) - \min_{a^j \in A^j} Q_i^{\pi^i, \pi^j}(s, a^j)
\]

Here, $Q_i^{\pi^i, \pi^j}(s, a^i)$ represents the expected return for agent $i$ under the joint policy $\pi^i, \pi^j$, while $\min_{a^j \in A^j} Q_i^{\pi^i, \pi^j}(s, a^j)$ represents the worst-case return if agent $j$ takes an adversarial action. In cooperative settings, the joint policy $\pi$ differs from adversarial policies, stabilizing the estimation of power.

To incorporate power into the learning process, the state-value function for agent $i$ is modified to include a power regularization term:

\[
V_i(s, a) = V_i^\pi(s, a) + \lambda V_i^{\pi, \rho_{i:j}}(s, a)
\]

Here, $V_i^\pi(s, a)$ is the original state-value function of agent $i$, and $V_i^{\pi, \rho_{i:j}}(s, a)$ represents the power component, which penalizes states where agent $j$ exerts high influence over agent $i$. This regularization can be viewed as a form of reward shaping, where the power measure is used to guide agents toward policies that are less vulnerable to adversarial influence.


\section{Power Regularization Over Communication}

Learning communication in cooperative settings can lead to more efficient coordination and strong, mutually dependent relationships among agents. However, misaligned agents can exploit these dependencies through sensory manipulation over the communication medium. Given the potential misuse of the communication medium, it is important to address how much dependency an agent delegates to other agents through the communication channel or protocol. Furthermore, it is imperative to ask how much dependency an agent should delegate to the communication medium, regardless of who uses the communication medium. In our work, we train policies to be more resilient to misaligned communication and miscommunication through adversarial training. Adversarial training is the practice of incorporating a variety of adversarial experiences and adversarial communication into training. We propose Communicative Power Regularization (CPR) to improve robustness against adverse impacts from misaligned communication by incorporating adversarial messages during training. Unlike standard power, which keeps the agent state constant, the communication message affects the perceived agent state and, therefore, can be viewed as a form of state regularization where the proximity of other states consisting of adversarial messages affects its perceived utility, similar to that of stochastic transitions by an environment.

\subsection{Communicative Power Regularization (CPR)}
We define communicative power as the decomposition of power into two components: standard power and power of communication. Standard power is the power delegated to other agents without leveraging the communication channel or protocol. In contrast, the power of communication is the power delegated to other agents over the communication channel or protocol. The total power \(\rho_{ij}^{\text{CPR}}\) that agent \(j\) has over agent \(i\) is defined as the sum of these two components:

\[
\begin{aligned}
\rho_{ij}^{\text{CPR}}(\pi^i, \pi^j, s^i, m^j) = 
& \, \rho_{ij}^{\text{Standard}}(\pi^i, \pi^j, s^i) \\
& + \rho_{ij}^{\text{Communication}}(\pi^i, \pi^j, s^i, m^j)
\end{aligned}
\]

Here, \(\rho_{ij}^{\text{Standard}}(\pi^i, \pi^j, s^i)\) represents the standard power, which quantifies the influence agent \(j\) has over agent \(i\) through actions alone, and \(\rho_{ij}^{\text{Communication}}(\pi^i, \pi^j, s^i, m^j)\) represents the power of communication, which quantifies the influence agent \(j\) has over agent \(i\) through communication.

The communicative power \(\rho_{ij}^{\text{CPR}}\) is defined as:

\[
\begin{aligned}
\rho_{ij}^{\text{CPR}}(\pi^i, \pi^j, s^i, m^j) = 
& \, Q_i^{\pi^i, \pi^j}(s, m^j, a^i, s^{i'}, m^{j'}) \\
& - \min_{a^j \in A^j} Q_i^{\pi^i, \pi^j}(s^i, m_{\text{adv}}^j, a^i, s^{i'}, m^{j'})
\end{aligned}
\]

This measures the difference in agent \(i\)'s expected return when agent \(j\) takes an adversarial action and sends an adversarial message \(m_{\text{adv}}^j\) compared to when agent \(j\) follows the joint policy.

Standard power can directly regulate misaligned communication in scenarios where communications are considered actions in the action space. However, its effectiveness in regularizing state-related misaligned communication assumes that appropriate variance is introduced into training, such as simultaneously learning a communication policy with an environment policy. Communicative power incorporates adversarial messages, whether they are individually sent or aggregated. This is particularly important in cases where individual messages are not misaligned but the aggregation of messages is misaligned.

In the traditional approach, the methodology does not provide direct defense mechanisms against adversarial attacks that specifically exploit model parameterization. Adversaries perform adversarial attacks to craft and inject adversarial samples, which usually target a model’s parameterization (e.g., a neural network’s decision boundary). However, some state-actions performed by certain agent roles contribute more to the environment’s expected utility than others, which finite-budget adversaries often consider. To address these challenges, we propose a framework that explicitly accounts for the influence of communication on power dynamics, ensuring robustness against both misaligned communication and adversarial exploitation of model parameterization.

The subsequent expressions are adapted from Li and Dennis \citep{li2024benefitspowerregularizationcooperative} to align with our proposed setting. To incorporate power into the learning process, the state-value function for agent \(i\) is modified to include a power regularization term:

\[
V_i(s, a) = V_i^\pi(s, a) + \lambda V_i^{\pi, \rho_{ij}}(s, a)
\]

Here, \(V_i^\pi(s, a)\) is the original state-value function for the task, and \(V_i^{\pi, \rho_{ij}}(s, a)\) represents the power component, which penalizes states where other agents exert influence over agent \(i\) through both actions and communication. The parameter \(\lambda\) is a scalar that controls the degree of power regularization over the expected return.

 Another approach involves applying both standard power and power of communication separately to enable individualized penalization of states where other agents exert greater control and penalization for states where there is excessive reliance on communicated messages.

The power regularization term \(V_i^{\pi, \rho_{ij}}(s, a)\) is defined as the sum of power rewards \(R_i^{\text{power}}(s_t, \pi)\) over states starting from \(s\) reached by unrolling the policy \(\pi\):

\[
V_i^{\pi, \rho_{ij}}(s, a) = \sum_{t=0}^T R_i^{\text{power}}(s_t, \pi)
\]

In the 2-agent setting, the power reward \(R_i^{\text{power}}(s, \pi)\) is defined as:

\[
R_i^{\text{power}}(s, \pi) = -\rho_{ij}^{\text{CPR}}(\pi^i, \pi^j, s^i, m^j)
\]

This indicates that the power reward penalizes the influence agent \(j\) has over agent \(i\) through both actions and communication in the state \(s\). In settings with more than two agents, the power reward captures the strongest individual influence that any other agent $j$ exerts on agent $i$:

$$
R_i^{\text{power}}(s, \pi) = -\max_{j \neq i} \rho_{ij}^{\text{CPR}}(\pi^i, \pi^j, s^i, m^j)
$$

By applying a maximization, this formulation emphasizes the worst-case dependency, making the regularization more sensitive to the most dominant external influence.

Our definition of communicative power is to further specify how power is allocated in the presence of a communication medium. This is in contrast to standard power, which makes no distinction over how power is distributed over coordinating devices or mechanisms. It is within the designer's discretion whether communication is appropriate for a cooperative task.


\section{Experiment Results}

\begin{table}[b]
    \vspace{0cm}
    \hfill
    \begin{minipage}{0.48\textwidth}
        \centering
        \renewcommand{\arraystretch}{1.6}
        \resizebox{\textwidth}{!}{%
        \begin{tabular}{c|c|c|c}
        \multicolumn{1}{c|}{\# of agents} & \multicolumn{2}{c|}{cooperative agents' scores} & improvement (\%) \\
        \cline{2-3}
        & With CPR & Without CPR &  \\
        \hline
        $\left[1,5\right]^*$ & \textbf{257.74 (\(\mp\) 45.93)} & 218.82 (\(\mp\) 66.40) & 18 \\
        $\left[2,4\right]^*$ & \textbf{204.92 (\(\mp\) 59.84)} & 93.56 (\(\mp\) 77.53) & 119 \\
        $\left[3,3\right]^*$ & \textbf{188.85 (\(\mp\) 88.50} & 33.53 (\(\mp\) 39.69) & 463 \\ \hline
        $\left[6,30\right]$ & \textbf{294.65 (\(\mp\) 31.21)} & 285.40 (\(\mp\) 43.90) & 3\\
        \hline
        $\left[4,8\right]$ & \textbf{242.25 (\(\mp\) 53.18)} & 116.27 (\(\mp\) 83.21) & 108 \\
        $\left[8,16\right]$ & \textbf{262.29 (\(\mp\) 46.94)} & 134.22 (\(\mp\) 86.52) & 95 \\
        $\left[12,24\right]$ & \textbf{273.94 (\(\mp\) 39.33)} & 142.86 (\(\mp\) 89.00) & 92 \\ \hline
        $\left[18,18\right]$ & \textbf{232.66 (\(\mp\) 77.85)} & 61.89 (\(\mp\) 64.19) & 276
        \end{tabular}%
        }
        \caption{Grid Coverage, Cooperative agents' scores (mean (\(\mp\) std.) over 100 trials), comparing performance with and without CPR across various [adversarial, cooperative] agent compositions. Asterisked (*) configurations denote training and evaluation with the same number of agents; others are scaled in evaluation.}
        \label{tab:gc_results}
    \end{minipage}
\end{table}

\begin{figure*}[t]
    \centering
    \includegraphics[width=\textwidth]{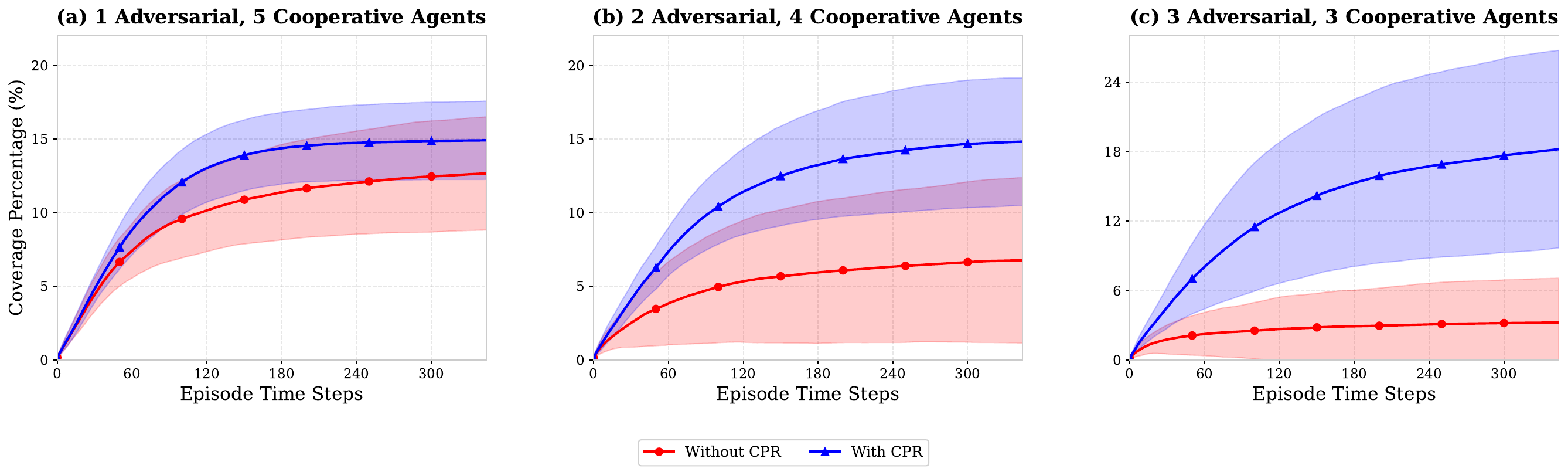}
    \caption{Grid Coverage, Average cooperative coverage percentage (over 100 trials) across varying team compositions, comparing agents trained with CPR (blue) vs without CPR (red).}
    \label{fig:coverage_15}
    \vspace{0.4cm}
\end{figure*}

We have three environments in which we evaluate CPR: (1) Grid Coverage \citep{blumenkamp2020emergenceadversarialcommunicationmultiagent} (GC), (2) Predator-Prey \citep{singh2018learning} (PP), and (3) A variant of Red-Door-Blue-Door (RDBD). We chose these settings because for CPR to be feasible in changing behaviors in the environment, we must have multiple, acceptable solutions that may be seen in cooperative, competitive, or mixed settings.

\textbf{Grid Coverage.} To evaluate the effectiveness of Communicative Power Regularization (CPR) in mitigating the effects of adversarial communication, we conducted experiments within the non-Convex Coverage map from the Adversarial Comms repository \citep{blumenkamp2020emergenceadversarialcommunicationmultiagent}. This environment challenges a team of cooperative agents to maximize area coverage on a grid map while contending with explicit adversarial agents designed to disrupt cooperative performance through the communication channel. Agents operate with limited local observations and communication ranges, utilizing a CNN-GNN-MLP architecture to process environmental input, exchange messages via the GNN, and select actions using the MLP.

Our evaluation directly compares the performance of cooperative MARL agents trained with CPR against baseline agents trained without CPR. Both sets of agents were evaluated over 100 trials in scenarios featuring varying numbers of cooperative and adversarial agents, where all agents, including adversaries, actively communicated throughout the episodes. This setup isolates the impact of CPR on the robustness of the cooperative strategy against communication-based attacks.

The comprehensive results presented in Table \ref{tab:gc_results} quantitatively demonstrate the significant advantage conferred by CPR across various team compositions and evaluation scales. We first established baseline performance in configurations where training and evaluation agent counts were identical ([1,5], [2,4], [3,3]). In these scenarios, cooperative agents employing CPR consistently achieved substantially higher scores, often with reduced variance as indicated by the standard deviations, compared to baseline agents without CPR. For instance, with 1 adversary and 5 cooperative agents ([1,5]), CPR yielded an 18\% improvement. This benefit became increasingly critical as the proportion of adversarial agents rose; in the challenging [3,3] scenario, CPR dramatically boosted the cooperative score, resulting in a 463\% improvement.
To assess the scalability of the learned policies, models trained on these starred configurations were then evaluated on significantly larger teams without retraining. Remarkably, CPR maintained substantial performance gains even in these scaled-up scenarios. For example, when scaling from the [2,4] training setup, CPR delivered improvements of 108\% for [4,8] and 95\% for [8,16]. Further scaling, such as to [12,24], still showed a 92\% improvement. Even in extremely scaled scenarios like [18,18] (scaled from [3,3]), CPR achieved a notable 276\% improvement. While the improvement in the [6,30] (scaled from [1,5]) configuration (3\%) was more modest, possibly due to the very low adversary-to-cooperative agent ratio. The overall trend strongly indicates that CPR facilitates the learning of robust coordination strategies that generalize effectively and preserve performance when deployed in larger, more complex multi-agent systems.

Figure \ref{fig:coverage_15} provides a more granular view, illustrating the average coverage percentage achieved by cooperative agents over episode time steps for [1,5], [2,4], and [3,3] configurations, respectively. In all depicted scenarios, the agents trained with CPR (blue curves) consistently outperform the baseline agents (red curves). They not only reach a higher final coverage percentage but also exhibit faster convergence towards their optimal performance early in the episode. Furthermore, the tighter variance bands (shaded regions) associated with the CPR agents suggest that CPR contributes to more stable and reliable performance across trials, reducing the detrimental impact of adversarial interference.

\begin{table}[b]
\renewcommand{\arraystretch}{1.2}
\begin{adjustbox}{width=\columnwidth,center}
\begin{tabular}{|c|c|}
\hline
    Description&Value \\
\hline 
    episode max timestep & 345 \\
\hline 
    communication range & 16 \\
\hline
    observation range & 8 \\
\hline
    \multirow{2}{*}{reward} & 
    +1 for visiting a previously \\
     & uncovered cell; 0 otherwise \\
\hline 
    agent actions & up, down, left, right, stay \\
\hline
    world shape & $24 \times 24$\\
\hline
    cooperative training & $20$ million time steps \\
\hline
    adversarial training & $20$ million time steps \\
\hline
    power regularization factor (\(\lambda\)) & $0.3$\\
\hline
    \end{tabular}
\end{adjustbox}
\caption{Grid Coverage, Experimental configuration parameters.}
\label{gc-config}
\end{table}

Synthesizing these findings, both the aggregate scores and the temporal coverage dynamics unequivocally show that CPR enhances the ability of cooperative agents to maintain effective coordination and achieve superior performance in the presence of adversarial communication. By regularizing the power or influence of messages, CPR fosters more robust communication strategies that are less susceptible to manipulation. This validation in the complex Grid Coverage task, featuring decentralized control and explicit adversaries, underscores the practical value of CPR for developing resilient multi-agent systems.

\begin{figure}[t]
    \centering
    \includegraphics[width=\linewidth]{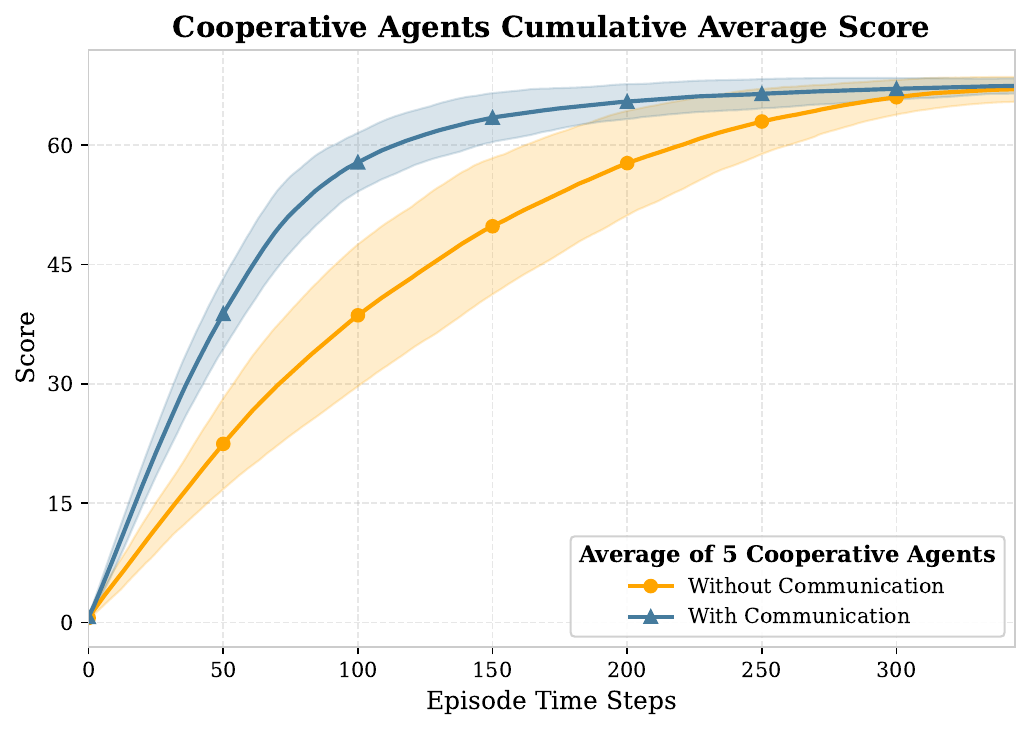}
    \caption{Grid Coverage, Cumulative average score of CPR-trained agents (5 cooperative, 100 trials), comparing performance with (blue triangles) and without (orange circles) communication}
    \label{fig:proof_of_comm}
    \vspace{0.7cm}
\end{figure}

An important concern is that CPR might inadvertently incentivize agents to avoid communication altogether to prevent penalties from adversarial messages. To investigate this, we conducted an ablation study and evaluated 5 CPR-trained cooperative agents, comparing their performance with active communication versus communication explicitly disabled. Figure \ref{fig:proof_of_comm}, showing cumulative average scores, clearly refutes this concern. Agents utilizing communication (blue triangles) significantly outperform the same agents operating without it (orange circles), achieving faster score accumulation and a higher final score. This confirms that CPR-trained agents do not abandon communication but learn to use it robustly, leveraging it for improved coordination and performance, thus demonstrating that CPR encourages resilient communication strategies rather than avoidance.

\textbf{Predator-Prey.}
The Predator-Prey environment is a grid-world scenario designed to evaluate MARL algorithms across cooperative, competitive, and mixed settings. In this setup, $N{-}1$ predators (specifically, 3 in our experiments) aim to capture a single prey agent. All agents possess limited local vision. During cooperative interactions, predators utilize communication to coordinate their approach to the prey's location. A time-step penalty encourages predators to find the most efficient path.

Our experimental configuration for Predator-Prey is detailed in Table \ref{pp-config}. The reward mechanism incorporates $\delta_i$, indicating whether agent $i$ has reached the prey; $n_t$, the count of agents at the prey's location at timestep $t$; and $\xi$, a parameter distinguishing competitive ($\xi = -1$), mixed ($\xi = 0$), and cooperative ($\xi = 1$) objectives. Each agent's observation space consists of a tensor reflecting its local visual field, employing one-hot encoding for the locations of predators and the prey.

We established a baseline using IC3Net trained under cooperative settings with communication always enabled (referred to as 'IC3Net(always-comm baseline)', see Table \ref{pp-results}). This baseline represents controllers reliant on continuous communication during cooperative training. Evaluating this model with communication disabled during testing ('IC3Net(no comm)' in Figure \ref{pp-results}) revealed a drop in performance (0.84 success rate vs. 1.0 during training), demonstrating that the cooperatively trained agents developed policies overly dependent on communication and lacked self-sufficiency when messages were absent. This finding highlights a potential vulnerability: even in cooperative scenarios without explicit adversaries, the unexpected absence of communication from a teammate can disrupt coordination, mimicking the effect of adversarial interference.

Subsequently, we trained an IC3Net model incorporating communicative power regularization ('IC3Net(CPR)') within the cooperative setting ($\lambda = 0.25$, Table \ref{pp-results}). Test results demonstrate that these agents maintained high performance even when communication was disabled ('IC3Net(CPR no comm)' in Table \ref{pp-results}, success rate 1.0), indicating improved robustness compared to the baseline.

We further investigated performance in a competitive setting. The baseline IC3Net model, trained cooperatively, failed entirely when tested in the competitive environment (success rate 0.0), despite communication channels remaining open ('IC3Net(always-comm)' competitive test, Table \ref{pp-results}). This underscores the difference between merely tolerating the absence of communication (where agents might learn redundant strategies) and possessing resilience against potentially uncooperative or misleading communication inherent in competitive settings. Policies that simply minimize communication among non-cooperative agents are not necessarily prepared for adversarial communication that exploits naive trust in received messages.

\begin{table}[t]
    \hfill
    \begin{minipage}{0.48\textwidth}
\centering
\renewcommand{\arraystretch}{1.4}
\resizebox{\textwidth}{!}{%
\begin{tabular}{c|c|c|c}
test/train & algorithm & setting & success \\ \hline
train & IC3Net(always-comm baseline) & cooperative & 1.0 \\
test & \textbf{IC3Net(no comm)} & cooperative & \textbf{0.84} \\
test & \textbf{IC3Net(always-comm)} & competitive & \textbf{0.0} \\ \hline
train & IC3Net(CPR always-comm) & cooperative & 1.0 \\
test & \textbf{IC3Net(CPR no comm)} & cooperative & \textbf{1.0} \\
test & \textbf{IC3Net(CPR always-comm)} & competitive & \textbf{1.0} \\
\end{tabular}
}
\caption{Predator-Prey, IC3Net success rates with/without CPR in cooperative/competitive settings and communication ablation tests.}
\label{pp-results}
\end{minipage}
\vspace{0.3cm}
\end{table}

\begin{table}[b]
\renewcommand{\arraystretch}{1.2}
\begin{adjustbox}{width=\columnwidth,center}
\begin{tabular}{|c|c|}
\hline
Description&Value \\
\hline
      \multirow{2}{*}{$\#$ of agents}& predator: 3\\
      & prey: 1\\
      \hline 
    \multirow{2}{*}{reward}  & $ r_i(t) = \delta_i * r_{explore}$ \\
    & $ + (1-\delta_i) * n_t^\xi * r_{prey} * \vert \xi \vert $ \\
    \hline 
     maximum timestep & 20 \\
     \hline
     predator actions & up, down, left, right, stay \\
     \hline
     observation &  (2*vision + 1$)^2$  \\
     space & $\times (\text{ohe-location, ohe-predator, ohe-prey})$\\
     \hline
     vision & 1 \\
     \hline
     grid-size& $5 \times 5$ \\
     \hline
     training& 2,000 epochs \\
     \hline
     testing& 1,000 episodes \\
     \hline
     power regularization factor (\(\lambda\)) & 0.25 \\
     \hline
\end{tabular}
\end{adjustbox}
\caption{Predator-Prey, Experimental configuration parameters.}
\label{pp-config}
\end{table}

\begin{table*}[t]
\centering
\renewcommand{\arraystretch}{1.4}
\resizebox{\textwidth}{!}{%
\begin{tabular}{c|c|c|c|c|c|c}
test/train & algorithm & setting & blue reward & red env reward & comm acc & episode len \\ \hline
train & MAPPO(cooperative baseline) & cooperative & 1.000 & 1.000 & 1.000 & 2.000 \\
train & \textbf{MAPPO(no adv-comm)} & competitive & 0.475 & -0.475 & 0.925 & 3.01 (\(\mp\) 0.005) \\
train & MAPPO(adv-comm (ideal)) & competitive & 0.499 (\(\mp\) 0.020) & -0.499 (\(\mp\) 0.020) & 0.411 & 3.000 \\
train & \textbf{MAPPO(CPR)} & cooperative & 0.464 & -0.464 & 1.0 & \textbf{3.000} (\(\mp\) 0.014) \\ \hline
test & \textbf{MAPPO(no CPR adv-comm)} & competitive & 0.368 & -0.368 & - & \textbf{3.64} (\(\mp\) 0.933) \\
test & \textbf{MAPPO(CPR adv-comm)} & competitive & 0.497 & -0.497 & - & \textbf{3.016} (\(\mp\) 0.127) \\
\end{tabular}
}
\caption{Red-Door-Blue-Door, MAPPO performance (rewards, episode length) with/without CPR under cooperative/adversarial communication.}
\label{rdbd-results}
\vspace{0.4cm}
\end{table*}

Conversely, the IC3Net model trained with communicative power regularization achieved successful task completion in the competitive setting, even with communication enabled ('IC3Net(CPR always-comm)' competitive test, Table \ref{pp-results}, success rate 1.0). This outcome suggests that applying power regularization over the communication channel guides agents towards learning more autonomous policies, equipping them to handle misaligned communication effectively.

More generally, our findings imply that incorporating CPR during training in cooperative settings can foster behaviors typically associated with policies learned under mixed or competitive conditions. The extent to which agent behavior deviates towards non-cooperative resilience depends on factors such as the power regularization factor ($\lambda$) and the permissible reduction in expected utility associated with adopting these more robust, autonomous strategies.

\begin{table}[b]
\renewcommand{\arraystretch}{1.2}
\begin{adjustbox}{width=\columnwidth,center}
\begin{tabular}{|c|c|}
\hline
Description&Value \\
\hline
      & blue: 1\\
      $\#$ of agents & red: 1\\
      & dummy-comm: 1 \\
      \hline 
      max timestep & 50 \\ 
      \hline
    reward & 
     $ r(s,a) =$
       $\begin{cases} 
  $(1/(t - 1)$ & $t $ < $ 50 \&\& \text{ all doors open}$ \\
  0 & t < 50 \\
  -1 & 50 \leq t \\
\end{cases} $   \\

\hline 
red reward& \multirow{2}{*}{- blue reward}\\
     (competitive)  &  \\
     \hline
     & door-status (red)\\
    observation & red-team status (dummy)\\
     & door-status + message (blue) \\
\hline
     training& 100,000 timesteps \\
\hline
     testing& 1,000 episodes \\
\hline
    power regularization factor (\(\lambda\)) & 0.75 \\
\hline
    \end{tabular}
\end{adjustbox}
\caption{Red-Door-Blue-Door, Experimental configuration parameters.}
\label{rdbd-config}
\end{table}

\textbf{Red-Door-Blue-Door}. To evaluate performance in a scenario demanding strict sequential coordination under potentially misleading communication, we utilize an adapted version of the Red-Door-Blue-Door (RDBD) environment \cite{lin2021learning}.
This version focuses solely on the coordination and communication aspects without grid-world movement, providing a clear testbed for communication robustness. The environment features three distinct policy roles: an Red agent, a Blue agent, and a Communication agent.

The Red and Blue agents are the primary actors. The Red agent can choose to open the red door or wait. The Blue agent can choose to open the blue door, open the red door, or wait. The cooperative task objective is achieved if the Red door is opened before or concurrently with the Blue door within 50 time steps, yielding a positive reward annealed by completion time. Opening the Blue door first or exceeding the time limit results in task failure and a negative reward. The Red agent possesses a hidden internal state determining whether its objectives are cooperative or adversarial (inducing a zero-sum reward structure between Red and Blue).

Communication is mediated by the dedicated Communication agent, which observes the Red agent's hidden state and generates a message $m \in \{0, 1\}$, intended to signal this state to the Blue agent at the start of each episode. The Blue agent observes the status of both doors and this message $m$, while the Red agent only observes door statuses. This setup explicitly isolates the communication channel. The challenge lies in the Blue agent learning appropriate reliance on $m$, as the Communication agent could represent a faulty, noisy, or explicitly adversarial channel sending misleading information. RDBD thus serves as a focused benchmark to assess if our proposed method (CPR) fosters policies robust against risks associated with the delegation of control via potentially compromised communication. Key configuration details are in Table \ref{rdbd-config}. 

Standard power and CPR are defined over Q-values estimations; however, for the implementation, we enact the immediate reward penalty through $k=1$-step adversarial action and adversarial message. This is defined in Equation \ref{equ}:

\begin{equation}
\label{equ}
    r_{i}(o,m_j,a_{i},a_{j}) = r_{i}(o,m_j,a_{i},a_{j}) + \lambda r_{i}(o,m_j^{adv},a_{i},a_j^{adv})
\end{equation}

Our experiments in RDBD (Table \ref{rdbd-results}) demonstrate CPR's effectiveness in enhancing robustness. The standard MAPPO baseline trained cooperatively (MAPPO(cooperative baseline)) learns an optimal, fast (ep. length 2.000), but communication-dependent strategy. Consequently, when faced with adversarial communication (MAPPO(no CPR adv-comm)), its performance collapses (blue reward 0.368 vs 1.0), highlighting the fragility of policies over-reliant on communication integrity.

In contrast, MAPPO agents trained with CPR ($\lambda=0.75$) exhibit significant resilience. While their cooperative strategy (MAPPO(CPR)) is slightly more cautious (ep. length $\approx$3.000, favouring the safer communication-independent sequence), they maintain strong performance under adversarial communication (MAPPO(CPR adv-comm)), achieving a blue reward of 0.497 and an episode length of 3.016. Notably, this performance is nearly identical to the ideal communication-independent strategy (MAPPO(adv-comm (ideal))).

This demonstrates that CPR successfully guides the Blue agent to mitigate risks by reducing dependence on the communication channel. By regularizing the influence exerted via communication during training, CPR encourages the agent to learn policies that are less susceptible to manipulation, effectively achieving robust, autonomous behavior akin to that learned under adversarial conditions, but within a cooperative training framework. These findings underscore CPR's value in building more reliable MARL systems by directly addressing vulnerabilities introduced through inter-agent communication.


\section{Conclusion}

This work tackled the inherent vulnerability of communicating MARL systems where reliance on information exchange can be exploited by misaligned agents. We argued that pervious work on power regularization, focused on action-based influence, inadequately captures the risks associated with delegating control through communication protocols. to address this, we introduced Communicative Power Regularization (CPR), a method to enhance MARL system robustness against misaligned communication by penalizing over-reliance on communication channels and increase self-autonomy of the agents. Extensive evaluations across Grid Coverage, Predator-Prey, and Red-Door-Blue-Door demonstrated CPR's ability to significantly improve cooperative performance and resilience in adversarial communication scenarios, without sacrificing communication when beneficial. While CPR involves a trade-off between robustness and optimal cooperative performance, it provides a practical framework for developing more secure and reliable cooperative MARL systems. Future work could focus on developing adaptive CPR frameworks, for instance by employing an adaptive Power Regularization Factor ($\lambda$), and on addressing heterogeneous trust dynamics. CPR offers a valuable step towards deploying resilient multi-agent systems in challenging real-world environments.



\begin{ack}
By using the \texttt{ack} environment to insert your (optional) 
acknowledgements, you can ensure that the text is suppressed whenever 
you use the \texttt{doubleblind} option. In the final version, 
acknowledgements may be included on the extra page intended for references.
\end{ack}



\bibliography{main}

\end{document}